\documentclass[english,epj]{svjour}
\usepackage[latin9]{inputenc}
\usepackage{amsmath}
\usepackage{graphicx}
\usepackage{amssymb}
\usepackage{esint}

\institute{Laboratoire PhLAM, UMR CNRS, CERLA, Université Lille 1, 59655 Villeneuve d'Ascq, France }
\abstract{The dynamics of cold atoms in conservative optical lattices obviously depends on the geometry of the lattice. But very similar lattices may lead to deeply different dynamics. In a 2D optical lattice with a square mesh, it is expected that the coupling between the degrees of freedom leads to chaotic motions. However, in somme conditions, chaos remains marginal. The aim of this paper is to understand the dynamical mechanisms inhibiting the appearance of chaos in such a case. As the quantum dynamics of a system is defined as a function of its classical dynamics -- e.g. quantum chaos is defined as the quantum regime of a system whose classical dynamics is chaotic -- we focus here on the dynamical regimes of classical atoms inside a well. We show that when chaos is inhibited, the motions in the two directions of space are frequency locked in most of the phase space, for most of the parameters of the lattice and atoms. This synchronization, not as strict as that of a dissipative system, is nevertheless a mechanism powerful enough to explain that chaos cannot appear in such conditions.
}

\usepackage{babel}

\begin{document}

\title{Synchronization in non dissipative optical lattices}

\author{D. Hennequin and P. Verkerk}

\maketitle

\section{Introduction}

Optical lattices are one of the most efficient tools to manipulate
cold atoms, by tuning or adjusting parameters such as the mesh and
height of the sites (atom confinement, atomic density), or the lattice
geometry. Thus it is not surprising that they became a toy model in
many fields. However, depending on the considered situation, the role
of the interactions between atoms varies a lot. Condensed matter systems
and strongly correlated cold atoms in optical lattices offer deep
similarities. The flexibility of the latter allowed the observation
of the superfluid-Mott insulator quantum phase transition \cite{mott},
of the Tonks-Girardeau regime \cite{tonks}, and more generally to
the superfluidity properties, including the instabilities. In these
experiments, the interactions between the cold atoms play a major
role, and require the use of a Bose-Einstein condensate. In particular,
instabilities are described by the Gross-Pitaevskii equation, through
the non-linear term \cite{kuan}. Quantum computing also required
a coupling between qubits. Optical lattices with controlled or {}``on
demand'' interactions appear to be an efficient implementation of
a Feynman's universal quantum simulator \cite{quantum simulator},
and are among the most promising candidates for the realization of
a quantum computer \cite{mandel2004}.

On the other hand, interesting behaviors can be found in noninteracting
systems. In many situations, including the one considered here, the
underlying physics is that of a single atom, without any interaction
between neighbors. The higher number of atoms simply increases the
observable signal. It is the case in statistical physics, where cold
atoms in optical lattices, through their tunability, made possible
the observation of the transition between Gaussian and power-law tail
distributions, in particular the Tsallis distributions \cite{tsallis,anders}.
Non-interacting cold atoms in optical lattices also allowed the observation
of Anderson localization \cite{anderson aspect,anderson roati,anderson garreau}.
Such cold atoms appear also to be an ideal model system to study the
dynamics in the classical and quantum limits. Both are closely related,
as the latter is only defined as a function of the former. For example,
quantum chaos is defined as the quantum regime of a system whose classical
dynamics is chaotic. A good understanding of the classical dynamics
is therefore an essential prerequisite to the study of quantum dynamics.
In non dissipative optical lattices, both the classical and the quantum
situations are experimentally accessible, and it is even possible
to change quasi continuously from a regime to the other \cite{raizen2000}.
Moreover, the extreme flexibility of the optical lattices makes it
possible to imagine a practically infinite number of configurations
by varying the complexity of the lattice and the degree of coupling
between the atoms and the lattice. Many results have been obtained
during these last years in the field of quantum chaos \cite{raizen2000,lignier2005}.
As mentioned above, these results in the quantum regime follow extensive
studies of the classical system \cite{CKR}.

Most of the above works used very simple potentials, mainly 1D. For
example, chaos is obtained only with a periodic (or quasi-periodic)
temporal forcing of the amplitude or frequency of the lattice \cite{raizen2000,lignier2005},
and only the temporal dynamics of the individual atoms is studied.
But recently, it appeared necessary to introduce more complex potentials,
in particular 2D potentials \cite{guo2009}. Although the dynamics
of particles in 2D potential has been extensively studied in the past,
it was mainly in model potentials \cite{chaikovsky1991,panoiu2000}.
Experimental optical lattices approach these models at best on a limited
domain, at the bottom of the wells. But in most cases, the potential
is more complex, and leads to a more complex and richer dynamics \cite{philo}.
Understanding acutely the classical dynamics of atoms in real potentials
is important, in particular because it has significant consequences
in the corresponding quantum systems.

The most common approach for the study of complex dynamics in conservative
systems is statistical, e.g. evaluating the percentage of the chaotic
area in the phase space. However, a more deterministic approach is
possible, as in dissipative systems. In a recent study, we studied
the dynamics of atoms in different 2D conservative optical lattices.
We chose experimentally feasible lattices, as these studies were motivated
by experiments, and we showed that different types of chaotic dynamics
appear, leading to different macroscopic behaviors: we showed in particular
that the lifetime of atoms in the lattices depend drastically on their
dynamics \cite{philo}.

One of the simplest experimental 2D conservative potential that we
studied in \cite{philo} is the square lattice, resulting from the
interference of 2 orthogonal pairs of counter-propagating stationary
waves. This lattice has a square mesh, and the two directions in space
are strongly coupled. Therefore the dynamics is expected to be fully
chaotic when anharmonicity is high enough, i.e. for high enough energy
of the atoms. This fully chaotic regime is effectively observed, except
when the lattice is red detuned. In this case, chaos disappeared almost
completely, and the dynamics remains essentially quasiperiodic, although
the nonlinearities remain the same. The reasons of this unusual behavior
was not discussed in\cite{philo}, where we focused on the chaotic
behaviors. However, the lack of chaos where it is expected deserves
to be studied in details, to understand what are the mechanisms inhibiting
its appearance?

In the present paper, we give some answers to this question by studying
in detail the square optical lattice with red detunings. We show that
at the bottom of the wells, the resonance frequencies in both directions
are degenerate, but when the atom energy increases, this degeneracy
should obviously disappear because of the anharmonicity of the potential.
However, we show that the motions in both directions remain locked
to the same frequency on a large domain, following a synchronization
mechanism close to the frequency locking process of dissipative systems.
Because of the conservation of energy, it is not a strict frequency
locking, but the quasiperiodic regime appears to be mainly a frequency
locked periodic regime with small sidebands. Even when the edges of
the wells are approached, chaos appears very marginally, in a regime
where the frequencies remain locked. The paper is organized as follows:
after this introduction, we discuss the model of the lattice, and
we search analytically for periodic solutions at the bottom of the
well, where approximations lead to a Duffing-like model. Then we discuss
the general case, i.e. far from the bottom of the well. In this case,
results are obtained mainly from numerical simulations. We show that
the main frequency of the motion is the same in both directions, whatever
the energy is, and we analyze the mechanisms leading to this synchronization.

\section{Description of the square lattice}

When cold atoms are dropped in a stationary wave, they undergo a force
$F$, the potential $U$ of which is proportional to the wave intensity
$I$, and inversely proportional to the detuning $\Delta$ between
the wave frequency and the atomic transition frequency:\begin{eqnarray*}
F & = & -\nabla U\\
U & \propto & \frac{I}{\Delta}\end{eqnarray*}
The detuning is a key parameter for the behavior of the atoms: they
accumulate in bright sites for red detunings ($\Delta<0$), and in
dark sites for blue detunings ($\Delta>0$). This paper is restricted
to the red detuned case, i.e. atoms located in the bright areas. When
the atoms are cooled with a Magneto Optical Trap (MOT), the atomic
density in these optical lattices is small enough to neglect the collisions
between atoms, and so the only source of dissipation is the spontaneous
emission. As spontaneous emission is proportional to $I/\Delta^{2}$,
it is relatively easy to build conservative optical lattices.

\begin{figure}
\includegraphics[bb=0bp 0bp 806bp 404bp,clip,width=8cm]{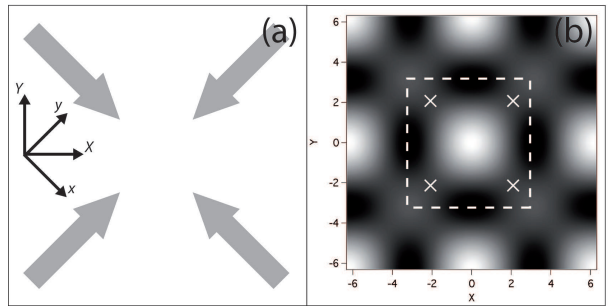}

\caption{\label{fig:fig1}a) Layout of the laser beams. b) Spatial distribution
of the intensity in the $\left(X,Y\right)$ space for $\alpha=0.5$.
Black corresponds to the minimum value (zero intensity), while white
corresponds to the maximum. The dashed square delimits the elementary
mesh of the lattice, and the white crosses are the saddle points. }

\end{figure}

The lattice geometry is a decisive parameter for the motion of the
atoms. Many geometries can be experimentally built, as a vertical
stack of ring traps \cite{courtade2006}, five-fold symmetric lattice
\cite{guidoni1999} or quasiperiodic lattices \cite{guidoni1997}.
But one of the simplest experimental optical lattice is the case of
two orthogonal stationary plane waves with the same linear polarization.
The configuration of the laser beams is shown on Fig. \ref{fig:fig1}a.
The total field is $\mathcal{E}=\cos kx+e^{i\phi}\sin ky$, where
$x$ and $y$ are the two space coordinates, $\phi$ a phase, $k=2\pi/\lambda$
the wave vector and $\lambda$ the wavelength of the laser beam. The
intensity can be written as:\[
I=\cos^{2}kx+\cos^{2}ky+2\alpha\cos kx\cos ky\]
where $\alpha=\cos\phi$. With the adequate normalization, the potential
is, for $\Delta<0$:\[
U=-I\]
When $\alpha=0$, the coupling between $x$ and $y$ disappears, and
the problem becomes separable. In all the other cases, the coupling
between $x$ and $y$ could induce complex dynamics. It is easy to
see that in these cases the elementary mesh of the potential is turned
of $\pi/4$ as compared to the $\left(x,y\right)$ axes, and thus
it is natural to introduce the following new coordinates: \begin{eqnarray*}
X & = & kx+ky\\
Y & = & ky-kx\end{eqnarray*}
The intensity and the potential can now be written:\begin{equation}
I=-U=1+\alpha\left(\cos X+\cos Y\right)+\cos X\cos Y\label{eq:potentiel}\end{equation}
The minus sign is that of the red detuning ($\Delta<0$) that we consider
here. In this conditions, the atoms accumulate in the bright areas:
the maxima of intensity correspond to the minima of the potential.
For blue detunings, atoms accumulate in the dark areas, leading to
significant differences in the motion \cite{philo}.

As an example, the following of the paper will be illustrated with
numerical values and plots obtained with $\alpha=0.5$. The choice
of this value is not restrictive, and these illustrations are representative
of the behaviors for other values of $\alpha$. Fig. \ref{fig:fig1}b
shows the spatial distribution of the intensity in this case. The
elementary mesh is indicated through the dashed line. Assuming $\alpha>0$,
the potential $U$ has its main well at the absolute minimum $E_{0}=-2\left(1+\alpha\right)$
at coordinates $\left(n2\pi,m2\pi\right)$, where $m$ and $n$ are
integers. It has also a relative minimum $-2\left(1-\alpha\right)$
in $\left(\pi+n2\pi,\pi+m2\pi\right)$. $\alpha=0$ is a special case
because the absolute and relative minima have the same value. The
intensity goes to zero in $\left(\pi+n2\pi,m2\pi\right)$ and $\left(n2\pi,\pi+m2\pi\right)$,
corresponding to the maxima of the potential. Two neighboring maxima
are separated by a saddle point where $U=E_{T}=-\left(1-\alpha^{2}\right)$
(remind that $\alpha\leq1$). $E_{T}$ is the minimum energy required
for a classical atom to jump from one well to another one. In the
quantum world, atoms can also tunnel through the barrier, leading
to a band structure. However, we do not have to consider the tunnelling
here, because, prior to any quantum treatment, the classical behavior
has to be deeply understood. On the other hand, it has been shown
that, for wells deep enough, the tunnelling vanishes \cite{bloch}.
Thus atoms, the energy $E$ of which is smaller than the threshold
$E_{T}$, are trapped into one site. On the contrary, atoms with $E>E_{T}$
can travel between sites, if they move in the right direction. It
is also important to note that these saddle points are on the bissectors,
connecting on a straight line a main well to a secondary one and again
to the next main well. Thus an atom with $E>E_{T}$ following this
straight line does not meet any obstacle: bissectors are clearly escape
lines.

Inside a trap site, the energy of the atom plays the role of a stochastic
parameter. Indeed, for low energies, the atoms remain located close
to the bottom of the well, and their dynamics can be approximated
by an harmonic motion. As the energy increases, the potential becomes
more and more anharmonic, the nonlinearities increase, and the dynamics
can become more and more complex \cite{philo}.

\section{The lattice around the origin: periodic solutions}

We study in the following the classical dynamics of cold atoms in
an optical lattice, and more generally that of a classical particle
in the corresponding potential. Let us first examine what is the motion
at the bottom of a main well. We approximate the potential to the
fourth order:

\begin{equation}
U=-\left(\alpha+1\right)\left(D_{X}+D_{Y}\right)-\frac{X^{2}Y^{2}}{4}\label{eq:duffing}\end{equation}
with

\begin{eqnarray*}
D_{X} & = & 1-\frac{X^{2}}{2}+\frac{X^{4}}{24}\\
D_{Y} & = & 1-\frac{Y^{2}}{2}+\frac{Y^{4}}{24}\end{eqnarray*}
$D_{X}$ and $D_{Y}$ have the shape of the potential of a conservative
Duffing oscillator. Thus the motion of the particles at the bottom
of a well appears to follow the dynamics of two coupled Duffing oscillator,
and so we can search for approximate harmonic solutions. The equations
to solve are immediately derived from eq. (\ref{eq:duffing}):\begin{subequations}\label{eq:acc}\begin{eqnarray}
F_{X} & = & -\left(\alpha+1\right)\left(X-\frac{X^{3}}{6}\right)+\frac{XY^{2}}{2}\label{eq:xacc1}\\
 & = & \ddot{X}\\
F_{Y} & = & -\left(\alpha+1\right)\left(Y-\frac{Y^{3}}{6}\right)+\frac{X^{2}Y}{2}\label{eq:yacc1}\\
 & = & \ddot{Y}\end{eqnarray}
\end{subequations}Let us first search if periodic oscillations are
approximate solutions of these equations. We search for:\begin{subequations}\label{eq:sol}

\begin{eqnarray}
X & = & X_{0}\cos\left(\omega t+\psi_{X}\right)\label{eq:xsol}\\
Y & = & Y_{0}\cos\left(\omega t+\psi_{Y}\right)\label{eq:ysol}\end{eqnarray}
\end{subequations}where $X_{0}$, $Y_{0}$, $\omega$, $\psi_{X}$
and $\psi_{Y}$ are constant. First, we look at the constraints that
have to be satisfied for the solutions. Then, to fully characterize
a solution, we need to check its stability, i.e. the behavior of the
trajectories close to that solution. In a dissipative system, a stable
periodic orbit is an attractor, and thus it plays a crucial role in
the effective dynamics of the system. On the contrary, an unstable
periodic orbit is not directly observable in experiments. The situation
is not so different in a conservative system. A periodic orbit is
stable if the behavior in its vicinity changes slowly with the distance,
i.e. if the trajectories change continuously from the periodic orbit
to a torus. The motion in the vicinity of the periodic orbit is then
not very different from a periodic oscillation. On the contrary, if
the periodic orbit is unstable, a small change in the initial conditions
leads to a completely different behavior, which cannot be considered
as a small perturbation of the initial periodic cycle.

The force terms in (\ref{eq:acc}) are developed in their Fourier
components, dropping the higher harmonics:\begin{subequations}\begin{eqnarray}
F_{X} & = & A_{X}\cos\left(\omega t+\psi_{X}\right)+B_{X}\sin\left(\omega t+\psi_{X}\right)\\
F_{Y} & = & A_{Y}\cos\left(\omega t+\psi_{Y}\right)+B_{Y}\sin\left(\omega t+\psi_{Y}\right)\end{eqnarray}
\end{subequations}$A_{X,Y}$ and $B_{X,Y}$ are the corresponding
Fourier components:\begin{subequations}\begin{eqnarray}
A_{X,Y} & = & \frac{\omega}{\pi}\int_{0}^{2\pi/\omega}dt\, F_{X,Y\,}\cos\left(\omega t+\psi_{X,Y}\right)\\
B_{X,Y} & = & \frac{\omega}{\pi}\int_{0}^{2\pi/\omega}dt\, F_{X,Y\,}\sin\left(\omega t+\psi_{X,Y}\right)\end{eqnarray}
\end{subequations}which lead to:\begin{subequations}\begin{eqnarray}
A_{X} & = & -X_{0}\left(1+\alpha\right)+\frac{X_{0}^{3}}{8}\left(1+\alpha\right)\nonumber \\
 &  & +\frac{X_{0}Y_{0}^{2}}{8}\left(2+\cos\left(2\psi_{Y}-2\psi_{X}\right)\right)\label{eq:ax}\\
B_{X} & = & -\frac{X_{0}Y_{0}^{2}}{8}\sin\left(2\psi_{Y}-2\psi_{X}\right)\\
A_{Y} & = & -Y_{0}\left(1+\alpha\right)+\frac{Y_{0}^{3}}{8}\left(1+\alpha\right)\nonumber \\
 &  & +\frac{X_{0}^{2}Y_{0}}{8}\left(2+\cos\left(2\psi_{Y}-2\psi_{X}\right)\right)\label{eq:ay}\\
B_{Y} & = & -\frac{X_{0}^{2}Y_{0}}{8}\sin\left(2\psi_{Y}-2\psi_{X}\right)\end{eqnarray}
\end{subequations}Differentiating twice the equations (\ref{eq:sol}),
we obtain another expression for the system equations which imposes
$B_{X}=B_{Y}=0.$ There are 6 periodic solutions for Eqs (\ref{eq:acc}):\begin{subequations}\begin{eqnarray}
X_{0} & = & 0\label{eq:sol1}\\
Y_{0} & = & 0\label{eq:sol2}\\
\psi_{X} & = & \psi_{Y}\label{eq:sol3}\\
\psi_{X} & = & \psi_{Y}+\pi\label{eq:sol4}\\
\psi_{X} & = & \psi_{Y}-\frac{\pi}{2}\label{eq:sol5}\\
\psi_{X} & = & \psi_{Y}+\frac{\pi}{2}\label{eq:sol6}\end{eqnarray}
\end{subequations}

The trivial solutions (\ref{eq:sol1}) and (\ref{eq:sol2}) are those
of a particle oscillating along one of the main directions, where
the minimum of the motion coincides with the bottom of the well. As
the potential is invariant when $X$ and $Y$ are exchanged, we have
to study only one of these solutions. The antiphase solution (\ref{eq:sol4})
corresponds to a change of sign of $Y$ as compared to in-phase solution
(\ref{eq:sol3}). As the potential is even in $X$ and $Y$, we need
to study only one of them. For the same reasons, only one of the quadrature
solutions (\ref{eq:sol5}) and (\ref{eq:sol6}) has to be studied.

\subsubsection*{Trivial solutions}

Let us first look at the $Y_{0}=0$ trivial solution. It comes immediately
from Eqs (\ref{eq:ax}) and (\ref{eq:acc}) that the frequency of
the motion is:\begin{equation}
\omega^{2}=\omega_{0}^{2}\,\left(1-\frac{X_{0}^{2}}{8}\right)\end{equation}
where $\omega_{0}=\sqrt{1+\alpha}$ is the frequency of the oscillations
at the very bottom of the wells. This relation is nothing but the
well-known dependence on the amplitude of motion, of the period of
oscillation for a simple pendulum $T=T_{0}\,\left(1+\frac{X_{0}^{2}}{16}\right)$.

Let us look in the neighboring of our trivial solution. For $Y$ small,
Eq. (\ref{eq:yacc1}) writes:\begin{equation}
\ddot{Y}+\omega_{0}^{2}\, Y-\frac{X^{2}Y}{2}=0\end{equation}
With a solution of the type of (\ref{eq:xsol}), this equation becomes\begin{equation}
\ddot{Y}+\left(\omega^{2}-\frac{X_{0}^{2}}{8}\left(1-\alpha\right)-\frac{X_{0}^{2}}{4}\,\cos2\omega t\right)Y=0\end{equation}
This equation is that of a parametric oscillator with a frequency
close to $\omega$, with no damping, and excited at the frequency
$2\omega$. Such an oscillator is known to diverge rapidly, and thus
the behavior in the vicinity of the trivial solutions does not remain
close to these solutions. Thus trivial solutions are unstable solutions
of the system.

\begin{figure}
\includegraphics[width=8cm]{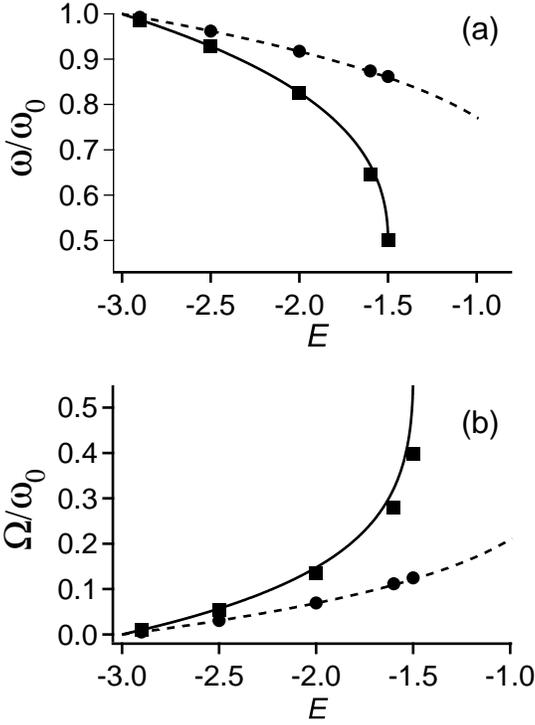}

\caption{\label{fig:freqs_duffing}Oscillation frequencies of the particle
in the vicinity of the periodic orbits, in the case of the Duffing
approximation. In (a), frequency of the periodic solution; in (b),
frequency of the beating. The solid (resp. dashed) lines refer to
the analytical in-phase (resp. quadrature) frequencies. The square
(resp. round) markers refer to the values computed through numerical
simulations, of the in-phase (resp. quadrature) solutions.}

\end{figure}

\subsubsection*{In-phase and antiphase solutions}

Let us now look at the (\ref{eq:sol3}) in-phase solution. Eqs (\ref{eq:ax})
and (\ref{eq:ay}) give two expressions for $\omega$, which are compatible
only if the motions in the two directions have the same amplitude
$X_{0}^{2}=Y_{0}^{2}$, with a frequency given by:\begin{eqnarray}
\omega_{1}^{2} & = & \omega_{0}^{2}-\left(1+\frac{\alpha}{4}\right)\frac{X_{0}^{2}}{2}\label{eq:freq_en_phase}\end{eqnarray}
This solution appears to be very similar to the trivial solutions:
the particles oscillate along a straight line, namely one of the bissectors.
Their frequency decreases as the motion amplitude increases. However,
as the stochastic parameter of our potential is the energy, it is
more convenient to represent the evolution of the dynamics as a function
of this parameter. The conversion between amplitude and energy depends
on the motion. For the present motion, the total energy $E_{1}$ leading
to a motion amplitude of $X_{0}$ is equal to the potential energy
in $\left(X_{0},X_{0}\right)$. Thus the relation between $X_{0}$
and $E_{1}$ is easily deduced from Eq. (\ref{eq:duffing}):\begin{eqnarray}
E_{1} & = & \omega_{0}^{2}\,\left(X_{0}^{2}-2\right)-\left(\alpha+4\right)\frac{X_{0}^{4}}{12}\label{eq:EvsX0}\end{eqnarray}

Fig. \ref{fig:freqs_duffing}a shows (solid line) the evolution of
$\omega_{1}$ as a function of the energy, for $\alpha=0.5$. In this
case, the bottom of the main well corresponds to an energy $E_{0}=-3$,
while the saddle in the full model corresponds to $E_{T}=-0.75$.
The Duffing model is an approximation of the motion at the bottom
of well, i.e. for energies of the order of $E_{0}$. For higher energies,
the Duffing approximation diverges from the full model. In particular,
the threshold energy in the Duffing approximation is $E_{D}=-1.5$,
and the in-phase solution vanishes for larger energies, as shown in
Fig. \ref{fig:freqs_duffing}a. The reason is that the motion occurs
exactly along the bissector, where is also located the threshold between
the wells. When the threshold energy is reached, the particle leaves
the well.

To estimate more precisely the domain of validity of the Duffing approximation,
it is more intuitive to use the motion amplitude rather than the energy.
Eq. (\ref{eq:EvsX0}) gives the relation between $E$ and $X_{0}$
for the (\ref{eq:sol3}) and (\ref{eq:sol4}) solutions: one finds
that $X_{0}=1$ corresponds to $E=-1.875$, which corresponds to the
half height of the non-approximated well. This means that the interval
on which the Duffing approximation is realistic is typically $-3<E\lesssim-1.875$.

To evaluate the stability of this solution, we study the motion in
its vicinity. We search for solutions of the type:\begin{eqnarray*}
X & = & \left(X_{0}+\varepsilon_{X}\right)\cos\left(\omega_{1}t+\phi_{X}\right)\\
Y & = & \left(X_{0}+\varepsilon_{Y}\right)\cos\left(\omega_{1}t+\phi_{Y}\right)\end{eqnarray*}
If this solution is injected into the motion equations, one finds
that $\varepsilon_{X}+\varepsilon_{Y}=C$ where $C$ is a constant.
It is easy to show that taking $C\neq0$ is equivalent to translate
from the solution $\left(X_{0},X_{0}\right)$ to the solution $\left(X_{0}+\frac{C}{2},X_{0}+\frac{C}{2}\right)$.
Thus we choose $\varepsilon_{X}=-\varepsilon_{Y}=\varepsilon$. The
equations of motion, in the slow varying envelop approximation (SVEA),
become:\begin{subequations}\begin{eqnarray}
\dot{\phi}_{X}-\dot{\phi}_{Y} & = & \left(1-\frac{\alpha}{2}\right)\frac{X_{0}}{2\omega_{1}}\varepsilon\label{eq:psi_dot}\\
\dot{\varepsilon} & = & -\frac{X_{0}^{3}}{8\omega_{1}}\left(\phi_{X}-\phi_{Y}\right)\label{eq:eps_dot}\end{eqnarray}
\end{subequations}or:\begin{eqnarray*}
\ddot{\varepsilon}+\Omega_{1}^{2}\varepsilon & = & 0\end{eqnarray*}
with\begin{eqnarray}
\Omega_{1} & = & \frac{X_{0}^{2}}{4\omega_{1}}\sqrt{1-\frac{\alpha}{2}}\end{eqnarray}
The perturbation of the in-phase solution oscillates, and thus the
in-phase solution is a stable periodic orbit. The amplitudes of the
motions along $X$ and $Y$ are modulated in antiphase at the frequency
$\Omega_{1}$. A phase modulation, in quadrature with the amplitude
modulation, is also present, as expressed in (\ref{eq:eps_dot}).
The neighboring solutions are quasiperiodic solutions with the same
main frequency $\omega_{1}$, and sidebands at the frequency $\Omega_{1}$
increasing with the energy (Fig. \ref{fig:freqs_duffing}b, solid
line). Note that for the highest energies ($E=-1.5$), $\Omega_{1}\simeq\omega_{1}$,
and thus the SVEA is no more valid. However, we already established
above that this energy domain is not fully relevant, as the Duffing
approximation is itself too much rough to describe the experimental
system for such energies. Thus the valid domain remains typically
$-3<E\lesssim-1.875$, as discussed previously.

As argued above, the $\psi_{X}=\psi_{Y}+\pi$ solution is the same
solution as $\psi_{X}=\psi_{Y}$ when $Y$ is changed in $-Y$. It
consists in a particle oscillating on the other bissector. This solution
$\left(X_{0},-X_{0}\right)$ is stable, and neighboring solutions
are an oscillating motion around the main solution. The dependence
between the frequencies and the energy is the same as in the $\psi_{X}=\psi_{Y}$
solution.

\subsubsection*{Quadrature solutions}

The last pair of solutions are those where $X$ and $Y$ oscillate
in quadrature of phase. The same steps as for the $\psi_{X}=\psi_{Y}$
lead to the solutions:\begin{subequations}\begin{eqnarray}
X & = & X_{0}\cos\left(\omega_{2}t\right)\\
Y & = & X_{0}\cos\left(\omega_{2}t\pm\frac{\pi}{2}\right)\end{eqnarray}
\end{subequations}with:\begin{eqnarray}
\omega_{2}^{2} & = & \omega_{0}^{2}-\left(2+\alpha\right)\frac{X_{0}^{2}}{8}\end{eqnarray}
The motion of a particle is a circle around the origin. The only difference
between both solutions is the rotation direction. The analysis of
the motion in the neighboring of these solutions leads to an oscillating
motion with a frequency:\begin{eqnarray}
\Omega_{2} & = & \frac{X_{0}^{2}}{4\omega_{2}}\sqrt{\frac{\alpha}{2}}\end{eqnarray}
These solutions are stable periodic orbits, and the motion in the
neighboring is quasiperiodic, with a main frequency $\omega_{2}$
decreasing with the energy, and sidebands at a frequency $\Omega_{2}$
increasing with the energy (Fig. \ref{fig:freqs_duffing}, dashed
lines). Note that for a circular motion, the kinetic energy never
vanishes: this is why this motion still exists for total energies
larger than $-1.5$, as it does not reach the saddle point.

To summarize the above results, we found that the dynamics in the
bottom of the main well of the lattice is organized around two unstable
periodic orbits and four stable periodic orbits. As expected, the
dynamics in the vicinity of these stable periodic orbits follows KAM
tori. These quasiperiodic regimes can be described as a motion at
the same frequency as the periodic orbits, perturbed by sideband components.
Although this system consists in two perturbed pendula, the effects
of the small perturbation is outstanding: whereas uncoupled pendula
can oscillate with any relative phase, the coupling introduced here
allows only four relative phases, even for very small motions, where
the perturbation tends to zero. To evaluate the pertinence of the
above description, let us now switch to the results of numerical simulations.

\section{Numerical simulations in the Duffing approximation}

The aim of numerical simulations is to determine how the previous
results evolve far from the bottom of the potential wells. Let us
first examine the results of numerical simulations in the Duffing
approximation. The interest of such simulations is to give complementary
information as compared to the previous results, as the shape of the
$X$ and $Y$ time evolution, or the complexity of the spectra.

\begin{figure}
\includegraphics[width=8cm]{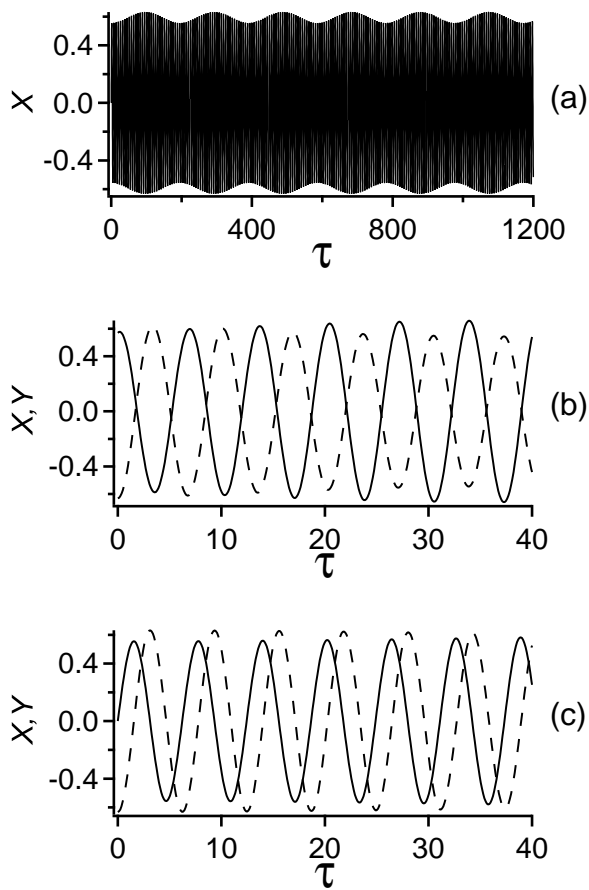}

\caption{\label{fig:signal}Motion of the particles as a function of the time
$\tau=\omega_{0}t$, for $\alpha=0.5$ and $E=-2.5$. (a) large scale
motion along $X$; (b) oscillations along $X$ (solid line) and $Y$
(dashed line) in the vicinity of the antiphase solution; (c) oscillations
along $X$ (solid line) and $Y$ (dashed line) in the vicinity of
a quadrature solution.}

\end{figure}

Fig. \ref{fig:signal} shows the time evolution of $X$ and $Y$ in
the vicinity of the antiphase solution (Fig. \ref{fig:signal}b) and
of one of the quadrature solution (Fig. \ref{fig:signal}c) for $\alpha=0.5$
and $E=-2.5$. In first approximation, it seems that the time evolution
in both directions follows the same frequency, with a clear phase
relation between them. However, the small amplitude modulation (Fig.
\ref{fig:signal}a) shows that other frequencies are present.

\begin{figure}
\includegraphics[width=8cm]{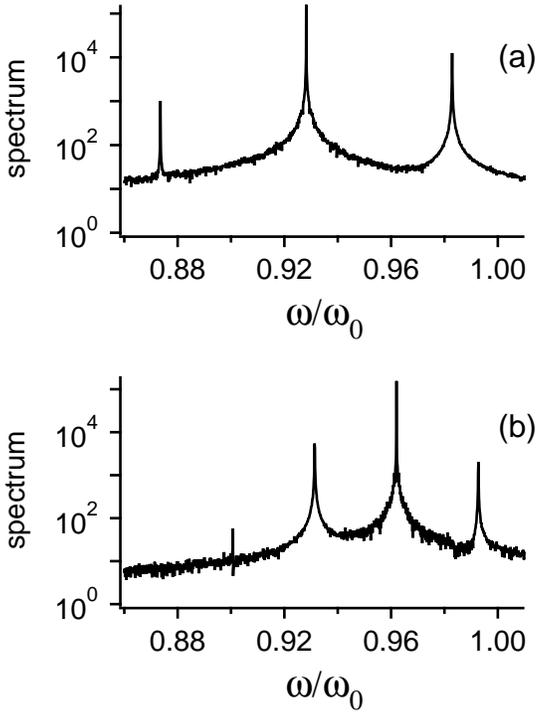}

\caption{\label{fig:fft}Spectrum of the motion of the particles along the
$X$ direction, in the vicinity of the (a) anti-phase solution, and
(b) quadrature solution. These are respectively the FFT of the signals
of Figs \ref{fig:signal}b and \ref{fig:signal}c. Both spectra are
represented with the same scales, to make the comparison easier.}

\end{figure}

More details can be found on the Fourier transform of the signals
(Fig. \ref{fig:fft}). Both dynamics appear to be driven by a main
frequency shifted as compared to the frequency $\omega_{0}=\sqrt{1+\alpha}$
at the bottom of the well. The sidebands are more than one order of
magnitude weaker, and the second sidebands are still two orders of
magnitude smaller. Thus the dynamical regimes along the $X$ and $Y$
directions can effectively be approximated as two frequency locked
oscillations, slightly perturbed by a sideband modulation.

It can be noticed that the spectra of Fig. \ref{fig:fft} are asymmetric.
On Fig. \ref{fig:fft}a, the high frequency sideband is larger than
the low frequency one, while it is inverted on Fig. \ref{fig:fft}b.
This sideband imbalance is simply due to the fact that we have both
amplitude modulation and phase modulation. The coupling between these
two modulations, expressed in (\ref{eq:psi_dot}) and (\ref{eq:eps_dot})
for the in-phase solution, leads to a high frequency component stronger
than the low frequency one, as observed in Fig. \ref{fig:fft}a. On
the contrary, for the quadrature solutions, the high frequency sideband
is predicted to be weaker than the low frequency one.

The frequencies obtained by the simulations are reported on Fig. \ref{fig:freqs_duffing}.
For low energies, they are identical to those predicted theoretically
in the previous section. For higher energies, small differences appear,
in particular for the sidebands of the antiphase solution. This originates,
as discussed previously, from the fact that the SVEA is no more valid
in this domain, as $\omega_{1}\simeq\Omega_{1}$.

\section{Behaviors far from the bottom of the wells}

In the above sections, the approximation of the optical lattice to
two coupled Duffing oscillators allows us to show that in the vicinity
of the bottom of the well, the dynamics is governed by a synchronization
process, very similar to the frequency locking. It is well known that
this type of phenomenon is able to inhibit complex dynamics in dissipative
systems, and if this synchronization applies to a large part of the
well, it could explain that chaos appears only into a very narrow
area. We study in this section the dynamics of the particles beyond
the domain where the Duffing approximation is valid.

Before to examine in details the dynamics of the particles in the
exact potential, let us remind the global evolution of the dynamics
in the Poincaré section. Our phase space is 4-dimensional, with directions
$\left(X,Y,\dot{X},\dot{Y}\right)$, but because of the energy conservation,
the accessible space reduces to a 3D surface. We choose to consider
Poincaré section at $\dot{Y}=0$ with increasing values, and thus,
Poincaré sections are in the 3D space $\left(X,Y,\dot{X}\right)$,
and they lie on a 2D surface $S_{P}$, which has the shape looking
like a semi-ellipsoid. To represent the Poincaré sections we could
project them on the $\left(X,Y\right)$ plane, but we use here the
more usual $\left(X,\dot{X}\right)$ plane. It shows the Poincaré
sections viewed from the vertex of the semi-ellipsoid. However , because
of the stiff sides of $S_{P}$, we have to keep in mind that many
curves are projected almost at the same location, and thus are superimposed.

\begin{figure}
\includegraphics[width=8cm]{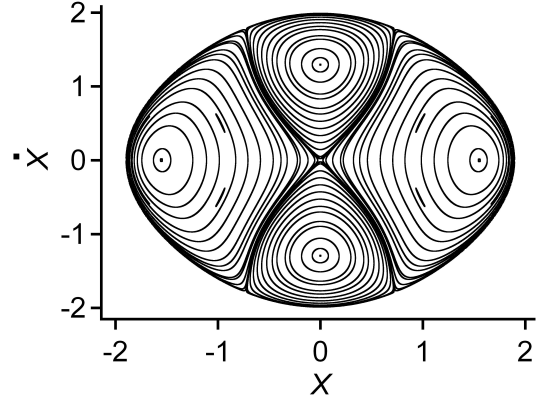}

\caption{\label{fig:ps}Poincaré sections for $\alpha=0.5$ and $E=-1.02$.}

\end{figure}

For relatively high energies, we know that the Duffing approximation
is no more valid. Thus it is interesting to look if, in spite of that,
the behaviors remain those described in the Duffing approximation.
As an example, we choose arbitrarily to illustrate the following of
the paper with the behaviors obtained for $E=-1.02$, i.e. an energy
well above the domain of validity of the Duffing approximation. These
behaviors are representative to those observed for all large energy
values, i.e. typically larger than $-2$. Fig. \ref{fig:ps} shows
the Poincaré section in this situation. These results have been obtained
through numerical resolution of the equations of motion which are
derived from the potential (\ref{eq:potentiel}), without any approximation.
We see four distinct domains separated by an X-shaped separatrix.
The central point $\left(X=0,\dot{X}=0\right)$ corresponds to the
first trivial solution (\ref{eq:sol1}). This solution has been found
to be unstable, which is compatible with the fact that all trajectories
move away from it. The second unstable periodic orbit corresponds
to $Y=0$, and so $\dot{Y}=0$, and thus it cannot be represented
in this Poincaré section through the $\dot{Y}=0$ plane.

\begin{figure}
\includegraphics[width=8cm]{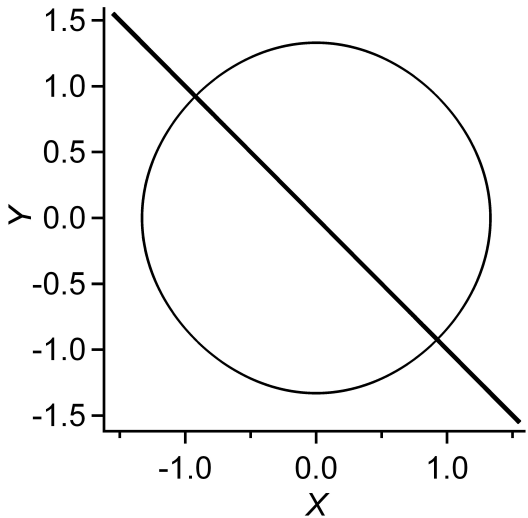}

\caption{\label{fig:solutions}Real space trajectories for the antiphase and
the quadrature solutions, calculated for $\alpha=0.5$, $E=-1.02$.
The antiphase solution consists in an oscillation of the particle
along the straight line, while the quadrature solutions correspond
to a rotation of the particle along the circle.}

\end{figure}

\begin{figure}
\includegraphics[width=8cm]{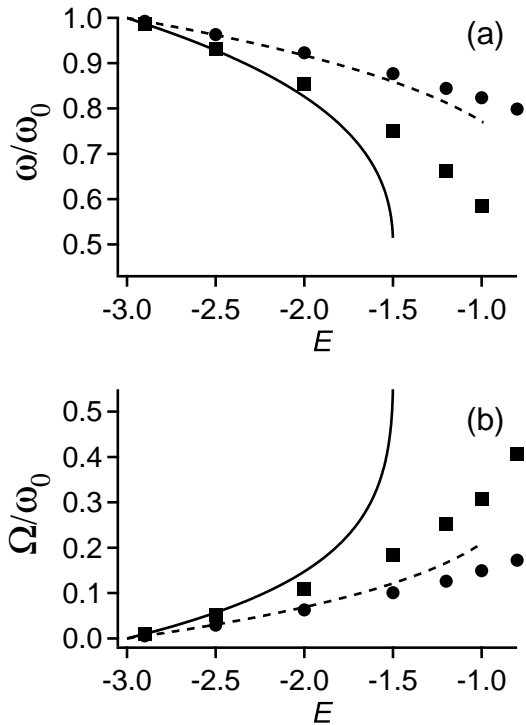}

\caption{\label{fig:freq_full}Oscillation frequencies of the particle in the
vicinity of the periodic orbits. In (a), frequency of the periodic
solution; in (b), frequency of the beating. The solid (resp. dashed)
lines are the same as in Fig. \ref{fig:freqs_duffing}, i.e. they
refer to the analytical in phase (resp. in quadrature) frequencies
in the Duffing approximation. The square (resp. round) markers refer
to the values computed, through numerical simulations of the full
model, for the in-phase (resp. in-quadrature) solutions.}

\end{figure}

In the right and left domains, the Poincaré sections are closed curves
around two points with $\dot{X}=0$. As by definition of the Poincaré
section, $\dot{Y}=0$, these points are turning points of the motion,
and thus correspond to the in-phase and antiphase solutions. In the
same way, in the top and bottom domains, the trajectories are also
cycling around two points, with $X=0$, which correspond to the quadrature
solutions. Fig. \ref{fig:solutions} shows the real space trajectories
of the antiphase solution and of one of the quadrature solutions.
The antiphase solution appears as a straight line on a bissector,
along which the particle oscillates with the frequency $\omega'_{1}$.
The quadrature solution appears as a circle around which the particle
turns with the frequency $\omega'_{2}$. The values of $\omega'_{1}$
and $\omega'_{2}$ are reported on Fig. \ref{fig:freq_full}a as a
function of the energy. As expected, the values predicted with the
Duffing approximation and those obtained from the full model diverge
significantly for energies larger than -2. However, the global evolution
remains that the main frequency decreases as energy is increased.

\begin{figure}
\includegraphics[width=8cm]{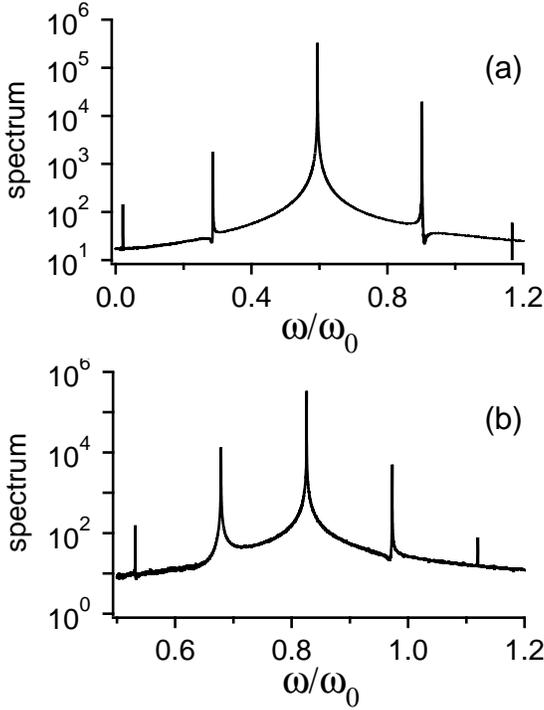}

\caption{\label{fig:fft_full}Spectrum of the motion along the $X$ direction,
in the vicinity of the (a) antiphase solution, and (b) quadrature
solution for $\alpha=0.5$, $E=-1.02$. The corresponding trajectories
give the Poincaré sections which are the first circles around the
periodic solutions in Fig. \ref{fig:ps}. In (a), Poincaré section
crossing the $\dot{X}=0$ axis in $X=1.46$. In (b), Poincaré section
crossing the $X=0$ axis in $\dot{X}=1.195$.}

\end{figure}

Let us now examine the dynamics on the tori close to a periodic stable
solution. We know that for lower energies, it can be considered as
periodic oscillations slightly perturbed by sidebands. To know if
this approximation is still valid for larger energies, we look at
the spectra of the trajectories. Fig. \ref{fig:fft_full} shows the
spectrum of two trajectories close to the antiphase and quadrature
solutions. The spectra have definitely the same characteristics as
those found for the lowest energies: the dynamics is mainly a periodic
oscillation, with the same frequency in both directions. This oscillation
is perturbed by sidebands, the amplitude of which is more than one
order of magnitude smaller. The second sideband is still one order
of magnitude smaller. Thus the behavior in the vicinity of the two
periodic orbits at large energy remains essentially a periodic oscillation,
where the motion along both directions is locked to the same frequency.
The frequency beating as a function of the energy is reported on Fig.
\ref{fig:freq_full}b: as for the main frequency, a clear divergence
as compared to the Dufffing approximation appears at energies larger
than -2, when the approximation is no more valid. But the global evolution
remains an increasing of the beating frequency with the energy.

\begin{figure}
\includegraphics[width=8cm]{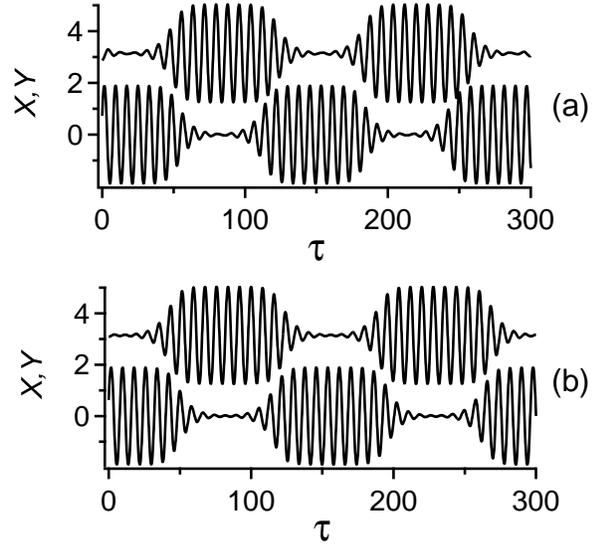}

\caption{\label{fig:unstable_signal}Motion of the particles as a function
of the time $\tau=\omega_{0}t$, for $\alpha=0.5$ and $E=-1.02$,
for initial conditions close to the unstable periodic orbits. In both
plots, $X$ and $Y$ trajectories are centered on $0$, but for sake
of clarity, the $Y$ trajectories have been shifted by $\pi$, so
that they appear to be centered on $\pi$. The corresponding trajectories
are on tori centered on (a) the antiphase periodic orbit and (b) a
quadrature periodic orbit.}

\end{figure}

\begin{figure}
\includegraphics[width=8cm]{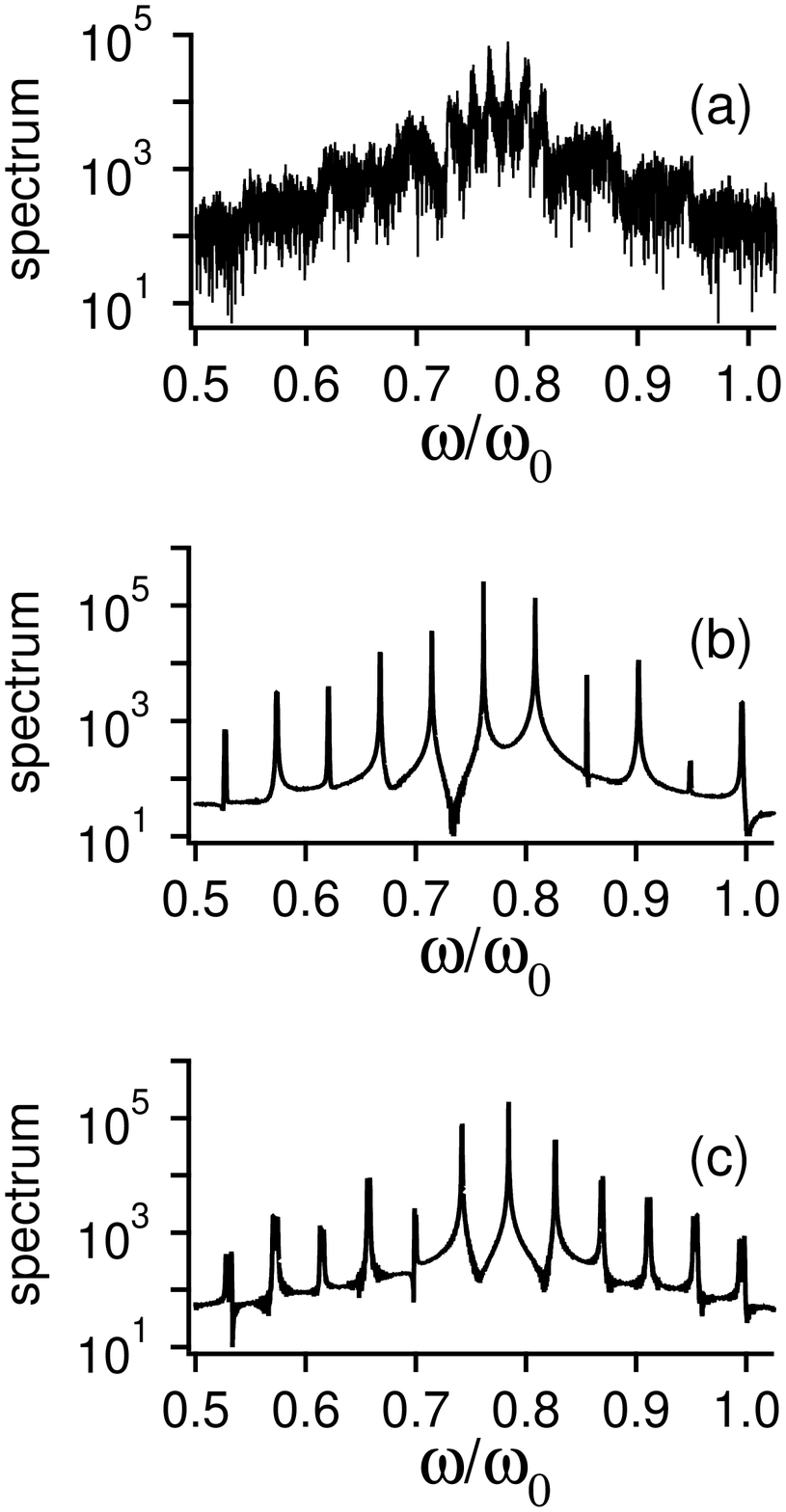}

\caption{\label{fig:unstable_fft}Spectrum of particle trajectories close to
the unstable periodic cycles, in (a) the chaotic area, (b) on the
side of the antiphase solution and (c) on the side of a quadrature
solution.}

\end{figure}

We examine now the behavior of the system in the vicinity of the unstable
solutions. As discussed in \cite{philo}, the dynamics in the immediate
vicinity of the separatrix is chaotic when the energy is large enough.
Fig. \ref{fig:unstable_fft}a shows a typical spectrum with the characteristics
of chaos, in particular a large continuous component. The existence
of chaotic trajectories in the vicinity of the trivial solutions confirms
the unstable nature of these solutions. However, the chaotic area
remains marginal in the lattice we discuss here. For example, in Fig.
\ref{fig:ps}, the chaotic area is so small that it cannot be visible
at the scale of the representation.

Fig. \ref{fig:unstable_signal} shows the temporal dynamics on tori
passing close to the ($X=0$, $\dot{X}=0$) point, just outside the
chaotic area, on the antiphase solution side in (a), and on the quadrature
side in (b). As expected in the vicinity of an unstable solution,
these dynamics do not seem to be linked to the solution itself. On
the contrary, they appear as an evolution of the regimes generated
by the corresponding stable periodic orbit. Indeed, they have the
main characteristics of the regimes observed in the vicinity of the
periodic cycles. In particular, the two regimes differ by the phase
difference between the oscillations in $X$ and $Y$, and the amplitude
modulations for the motion in $X$ and in $Y$ are always in antiphase,
i.e. the particle oscillates for a while along the $X$-axis and then
moves to the $Y$-axis, and vice-versa. Fig. \ref{fig:unstable_fft}b
and \ref{fig:unstable_fft}c shows the spectra associated with these
dynamics. The dynamics in the $X$ and $Y$ directions are still synchronized,
with one main frequency and sidebands well below this main frequency.
However, some differences appear as compared to the vicinity of the
stable solutions: in particular, the amplitude of the sidebands is
larger, with less than one order of magnitude as compared to the main
frequency, and several harmonics have a non negligible amplitude.

\begin{figure}
\includegraphics[width=8.5cm]{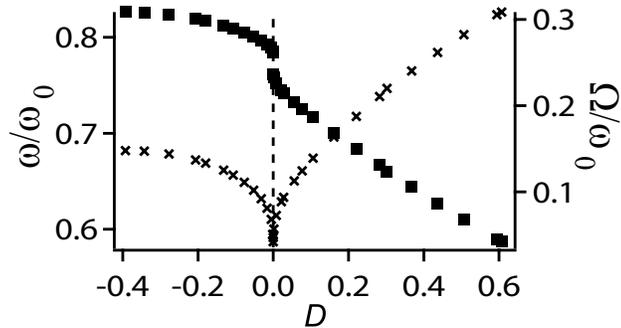}

\caption{\label{fig:freqs_sp}Evolution of the main frequency $\omega$ (squares)
and the beating $\Omega$ (crosses) for different trajectories, for
$\alpha=0.5$ and $E=-1.02$. Trajectories are identified through
the point where they intersect a line linking the quadrature and the
antiphase solution. $D$ is the distance between the intersection
point and the separatrix. The quadrature (resp. antiphase) solution
is in $D=-0.4$ (resp. $D=0.6$).}

\end{figure}

We have now a good picture of the motion of an atom in a well of our
lattice. Whatever its energy is, its motion is mainly a periodic oscillation
distorted by a slow drift. This drift vanishes on the periodic stable
solutions, and increases when one moves away from these solutions.
The frequency of the oscillation, together with that of the sidebands,
evolve with the energy, but evolve also in the phase space of a given
energy. Fig. \ref{fig:freqs_sp} shows the evolution of the behavior
when one moves away from the stable solutions, for $\alpha=0.5$ and
$E=-1.02$. On the whole accessible phase space, the dynamics is still
essentially an oscillation with the same frequency in both directions,
perturbed by sidebands. Fig. \ref{fig:freqs_sp} shows that the main
frequency evolves continuously between the resonance frequencies of
the different stable solutions. In particular, for the four domains
delimited by the separatrix, the main frequency tends to that of the
unstable solution for trajectories at the edge of the domain. The
beating frequency also evolves monotonically in each domain, tending
to zero when the separatrix is approached. So the transition between
each periodic solution occurs without any discontinuity.

\section{Conclusion}

We have shown in \cite{philo} that the nature and the complexity
of the motion of an atom in an optical lattice -- or similarly of
a classical particle in the corresponding external potential -- can
change drastically as a function of the lattice properties. We study
the fundamental mechanisms leading to so important differences in
the dynamics, and especially the absence of chaos in the potential
wells of a red detuned square optical lattice. We adopt an approach
quite unusual in the domain of conservative chaos, because our aim
is to find experimental tools able to characterize more precisely
the complex conservative dynamics, and in particular able to distinguish
between different complex behaviors. In square lattices, atoms traveling
between sites follow two different chaotic behavior, depending on
the lattice blue or red detuning. This difference probably originates
in the behavior of atoms inside the wells: full chaos appears when
the laser frequencies are blue detuned, whereas chaotic trajectories
are quasi inexistent when they are red detuned. The former is not
surprising: the edges of a well are far from harmonicity, and the
coupling between the two directions, together with the nonlinearities,
become very high. On the contrary, the mechanism leading to the absence
of chaos in the latter needed to be clarified. We show here that synchronization
between the motion in the two directions of space inhibits chaos,
through a mechanism very similar to that of frequency locking in dissipative
systems. Indeed, the frequencies in both directions of space are degenerate
at the bottom of the well. For atoms with higher energies, the degeneracy
should be broken, but the coupling overcomes the anharmonicity and
leads to a motion with one single frequency. Synchronization is not
strictly frequency locking, as our system is conservative, and there
is no dissipation to compensate for frequency pulling. So the periodic
oscillations are modulated in phase and in amplitude, i.e. the main
frequency comes along with small sidebands. In the phase space, the
amplitude modulation can increase, and even reach 100\% in the vicinity
of the unstable periodic orbits, but the description in terms of frequency
locking remains valid on most of the phase space. Finally, in our
conservative lattice, synchronization appears to be at the origin
of the absence of chaos, following the same mechanisms as in dissipative
systems. This intrinsic origin of this inhibition is not a local behavior,
i.e. the lack of chaos will remain whatever the lattice parameters
are, except for the lattice frequency. In particular, the quantum
dynamics should be studied in another lattice, as e.g. the blue detuned
square lattice, where chaos is fully developed.

This analysis shows that the dynamics of conservative systems can
be characterized using deterministic tools rather than the traditional
statistical approaches. Atoms in optical lattices are a good model
system to apply these tools and test them on experiments. Future studies
could refine the present results, in particular by checking that this
approach is still valid when, at the main resonance, the frequencies
in both spatial directions are not equal, but only multiple. This
can be achieved with the blue detuned square lattice, by choosing
adequately $\alpha.$

\end{document}